\documentclass[prx,twocolumn,english,superscriptaddress,floatfix,longbibliography]{revtex4-2}

\usepackage{hyperref}
\hypersetup{
    unicode=true, 
    colorlinks=true, 
    linkcolor=blue, 
    citecolor=blue
}

\usepackage{latexsym,amssymb,bm,bbold,amsfonts,amstext,graphicx,bbm,bm,relsize,dsfont,times}

\usepackage[table]{xcolor}
\usepackage{amsmath} 
\usepackage{amsthm} 
\usepackage{enumerate} 
\usepackage{float}

\makeatletter
\def\@bibdataout@aps{%
\immediate\write\@bibdataout{%
@CONTROL{%
apsrev42Control%
\longbibliography@sw{%
    ,author="08",editor="1",pages="1",title="1",year="1"%
    }{%
    ,author="08",editor="1",pages="1",title="1",year="1"%
    }%
  }%
}%
\if@filesw \immediate \write \@auxout {\string \citation {apsrev42Control}}\fi 
}
\makeatother
\definecolor{blue-violet}{HTML}{c13bff}

\usepackage{tikz}
\newcommand{\tr}{\mathrm{Tr}}

\def\bra#1{\langle{#1}|}
\def\ket#1{|{#1}\rangle}
\def\braket#1{\langle{#1}\rangle}
\newcommand{\ketbra}[2]{\ket{#1}\!\bra{#2}}

\def\BraVert{\egroup\,\mid\,\bgroup}

\newcommand{\ex}{\text{ex}}
\newcommand{\exph}{\text{ex-ph}}
\newcommand{\ph}{\text{ph}}

\newcommand{\F}{\text{F}}
\newcommand{\CT}{\text{CT}}

\usepackage{tcolorbox}

\definecolor{light_blue}{HTML}{f0f5ff}

\tcbset{fontupper=\normalsize,
colback=white, colframe=black, colbacktitle=light_blue,
coltitle= black, arc=0.5mm, center title}
\makeatletter
\def\l@subsubsection#1#2{}
\makeatother

\begin{document}

\title{Exciton transport in amorphous polymers 
and the role of morphology and thermalisation}
\author{Francesco Campaioli}
\email{francesco.campaioli@rmit.edu.au}
\affiliation{Chemical and Quantum Physics, and ARC Centre of Excellence in Exciton
Science, School of Science, RMIT University, Melbourne 3000, Australia}

\author{Jared H. Cole}
\email{jared.cole@rmit.edu.au}
\affiliation{Chemical and Quantum Physics, and ARC Centre of Excellence in Exciton Science, School of Science, RMIT University, Melbourne 3000, Australia}

\date{\today}

\begin{abstract}
\noindent
Understanding the transport mechanism of electronic excitations in conjugated polymers is key to advancing organic optoelectronic applications, such as solar cells, OLEDs and flexible electronics. While crystalline polymers can be studied using solid-state techniques based on lattice periodicity, the characterisation of amorphous polymers is hindered by an intermediate regime of disorder and the associated lack of symmetries. To overcome these hurdles we use a reduced state quantum master equation approach based on the Merrifield exciton formalism. Using this model we study exciton transport in conjugated polymers and its dependence on morphology and temperature. Exciton dynamics consists of a thermalisation process, whose features depend on the relative strength of thermal energy, electronic couplings and disorder, resulting in remarkably different transport regimes. By applying this method to representative systems based on poly(p-phenylene vinylene) (PPV) we obtain insight into the role of temperature and disorder on localisation, charge separation, non-equilibrium dynamics, and experimental accessibility of thermal equilibrium states of excitons in amorphous polymers.
\end{abstract}

\maketitle
\makeatletter

\section{Introduction}
\label{s:introduction}
\noindent
Organic semiconductors (OSCs) are at the forefront of the current research efforts for the development and improvement of optoelectronic technology, such as solar cells~\cite{Hoppe2008,Zhang2018,Wang2019b}, organic light-emitting diodes (OLEDs)~\cite{Zampetti2019,Manousiadis2020}, thin-film transistors~\cite{Salleo2002,Cho2008,Wu2018,Zou2019}, sensors and flexible electronics~\cite{Coropceanu2007,Mikhnenko2015,Kang2015,Gao2019}. In particular, conjugated polymers are a key class of OSCs for photovoltaic applications, due to their ability to transport electronic excitations, i.e., \emph{excitons}, over tens of nanometers, together with their low productions cost, ease of fabrication and flexibility~\cite{Brixner2017,Jang2018}. Depending on their chemical composition and fabrication conditions, polymeric semiconductors can be found in a variety of different morphologies characterised by specific exciton transport properties~\cite{Ostroverkhova2016}.

While the fundamental features of energy and charge transport across crystalline, semicrystalline and amorphous polymers are qualitatively known, the dependence of exciton dynamics on temperature and morphology is still an important area of investigation~\cite{Arago2015,Horak2018,Lyskov2019,Binder2020}. For example, the introduction of amorphous polymers characterised by high electron mobility has challenged the idea that crystalline ones would be optimal for charge transport~\cite{Collini2009,McCulloch2012,Nielsen2013,Qin2013,Groff2013,Ma2014}. 
The study of exciton transport in amorphous polymers is however hindered by the complexity that arises from their disordered nature, which prevents the use of solid-state techniques based on lattice periodicity. In this intermediate transport regime between band conduction and incoherent hopping, semiclassical techniques such as Marcus theory often fail to reproduce the coherent quantum attributes of the exciton dynamics~\cite{Yost2012,Stehr2014,Kranz2016a,Taylor2018}, while first-principles calculations like multiconfiguration time-dependent Hartree method (MCTDH) become computationally intractable due to the large size of the systems of interest~\cite{Binder2013,Oberhofer2017,Binder2018,Lyskov2019,Balzer2021,Binder2020}.

An effective alternative to study exciton transport in disordered OSCs is given by reduced state quantum master equations~\cite{Jang2018}. These allow for the efficient description of the interaction between excitons and nuclear vibrations, i.e., \textit{phonons}, and account for both coherent and incoherent dynamics. 
Quantum master equations have therefore been applied to a variety of exciton transport problems in OSCs, such as natural photosynthetic complexes~\cite{Mohseni2008,Rebentrost2009,Caruso2009,Caruso2010}, molecular aggregates~\cite{Ohta2006,Nakano2016,Nakano2019,Xie2019}, and disordered systems~\cite{Lee2015,Balzer2021}.
A key insight from this body of literature is that exciton-phonon interactions have an essential role for exciton transport in OSCs. On one hand, decoherence induced by local and uncorrelated phonon modes is known to improve transport efficiency in disordered systems~\cite{Plenio2008,Rebentrost2009,Kassal2013}, counteracting weak and strong localisation phenomena, such as Anderson localisation~\cite{Anderson1958a,Wang2006,Lagendijk2009,Walschaers2013}, that otherwise prevail when dynamics is primarily coherent. On the other hand, strong spatial correlations between bath modes can facilitate the emergence of decoherence-free subspaces, improving transport of certain states in ordered systems~\cite{Jeske2013}. 

Recently, quantum master equations have also been used to study exciton transport in conjugated polymers, albeit only for the case of crystalline morphologies. In Ref.~\cite{Lyskov2019}, Lyskov \emph{et al.} bridge first-principles calculations to a Lindblad master equation~\cite{Gorini1975,Lindblad1976,Lidar2001,Heinz-PeterBreuer2002,Milz2017}, and obtain a phenomenological model for triplet exciton dynamics in crystalline poly(p-phenylene vinylene) (PPV). Their results show that, conversely to the case of disordered OSCs, decoherence slows down exciton dynamics in crystalline PPV, inducing a rapid transition from ballistic (coherent) to diffusive (incoherent) transport. However, the model used in Ref.~\cite{Lyskov2019} cannot be directly applied to amorphous polymers, whose difficult theoretical characterisation presents several questions~\cite{Gowrishankar2006,Gowrishankar2008,Beenken2009,Nayyar2011,Kilina2013,Bhatta2014,Sajjad2015,Bredas2016,Simine2017,Binder2018,Noriega2018,Scholes2019,Pandya2020}. How do temperature and morphology affect exciton transport? What are the features of the dynamical transition between localised and delocalised states? Are thermal equilibrium states experimentally accessible? How does disorder affect charge-separation dynamics?

Here we answer these questions generalising the master equation approach used in Ref.~\cite{Lyskov2019} to conjugated polymers with arbitrary morphology. 
Firstly, since excited electronic states of amorphous polymers are known to display charge separation over a few monomers~\cite{Qin2013,Ma2014}, we formulate a master equation for the dynamics of Merrifield excitons~\cite{Merrifield1961,Binder2013}, i.e., a strongly coupled electron-hole pair. Accordingly, our approach can be seen as a direct extension of the master equations used in Refs.~\cite{Rebentrost2009,Caruso2009,Caruso2010,Lyskov2019} for the dynamics of charge-neutral Frenkel excitons~\cite{Frenkel1931}, characteristic of OSCs with low dielectric constant.
 
Furthermore, our model, introduced in Sec.~\ref{s:methodology}, provides a rigorous thermodynamic description of exciton-phonon interactions, overcoming the limitations of the phenomenological approach used in Ref.~\cite{Lyskov2019}. There, the authors fit the master equation parameters to MCTDH calculations carried out on small two-monomer subsystems over short time scales. 
Such approach, affected by artificially short recurrence times, may fail to correctly reproduce the dynamics for long time scales and returns decoherence rates that do not necessarily obey thermodynamics in the long-time limit. Here, instead, we use the Bloch-Redfield formalism to calculate temperature-dependent decoherence rates from the phonons' correlation functions~\cite{Lidar2001, Jeske2015}, to then express the dynamics of the exciton's reduced state using a master equation of the Lindblad form~\cite{Heinz-PeterBreuer2002,Mohseni2008,Nakano2016,Nakano2019,Xie2019}. This approach has the additional advantage of providing ensemble average transport properties without calculating many individual stochastic realisations of system-environment dynamics.

In Sec.~\ref{s:results}, we use the model to study the exciton transport properties of some representative oligo and poly(p-phenylene vinylene) (OPV and PPV) systems, across different morphologies, from crystalline to highly disordered. Instead of relying on material-specific modelling of polymers, we express our calculations in terms of electronic couplings and thermal energy. This allows us to explore the impact that disorder and temperature have on localisation, charge-separation dynamics and equilibration time scales. Exciton transport is therefore interpreted as a thermalisation process~\cite{Figueirido1992,Lidar2001,Heinz-PeterBreuer2002,Lidar2003, Jeske2015, Binder2015}, whose features vary with the relative strength between thermal energy, electronic couplings and disorder.
We show how such differences in electronic coupling determine whether the phonon bath \textit{cools} or \textit{heats} the system, with dramatic effects on exciton transport properties.
We conclude discussing our results in Sec.~\ref{s:conclusions}, where we compare and contrast our findings with those of other works, and outlining the use of our method to exciton transport problems of great importance for optoelectronic applications.

\section{Methodology}
\label{s:methodology}
\noindent
The exciton transport model that we introduce in this article is based on the following two premises. First, the characteristic times and energies associated with exciton transport in OSCs at room temperature and natural illumination imply that excitons do not interact with each other, as they are sufficiently spatially and temporally separated~\cite{Jang2018}. Under these circumstances, it is a standard approach to focus on the transport of a single electronic excitation, formally defined by the single-exciton manifold~\cite{Rebentrost2009,Caruso2009,Caruso2010,Binder2013,Jang2018,Lyskov2019,Binder2020}.
Second, we consider polymeric materials whose structural changes are far slower than the characteristic femtosecond to nanosecond time scales of exciton dynamics, such as for the paradigmatic case of PPV and its amorphous derivative MEH-PPV~\cite{Qin2013,Ma2014}. This allows us to adopt a snapshot approach for a medium's morphology, which is assumed to be time-independent. Additionally, we assume that the interaction between excitons and rapidly oscillating conformational coordinates like bond-length alternations, here modelled as a bath of independent harmonic oscillators, are well described in the weak coupling approximation~\cite{Mohseni2008,Jang2018}.

\begin{figure}[t]
    \centering
    \includegraphics[width=0.48\textwidth]{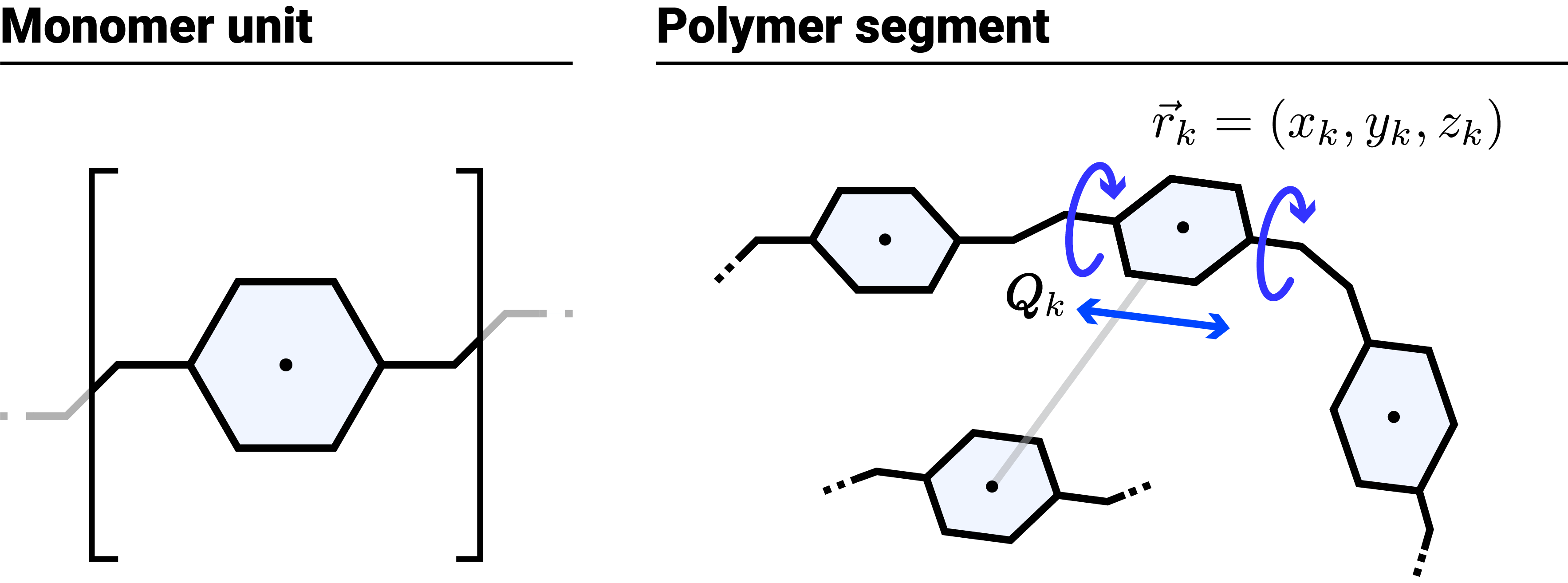}
    \caption{Polymeric materials are here modelled as a network of $N$ nodes, each of which represents a monomer. Monomers positions $\vec{r}_k$ and conformational coordinates $\bm{Q}_k$ define the morphology of the medium and the structure of the exciton's Hamiltonian. Exciton transport on such network is mediated by through-bond couplings between conjugated monomers that are part of the same polymer, or by through-space couplings (\textit{gray line}) between pairs of monomers in proximity of each other.}
    \label{fig:methodology}
\end{figure}

Under these assumptions, the transport of a single electronic excitation in OSCs is typically modelled as a \emph{quantum stochastic walk} (QSW) of a charge-neutral Frenkel exciton over a network of $N$ nodes~\cite{Whitfield2010,Sinayskiy2019}, each of which represents a monomer or a chromophore~\cite{Mohseni2008,Caruso2009,Jang2018,Lyskov2019}. However, motivated by recent results on the electronic excitations of amorphous polymers~\cite{Qin2013,Ma2014}, we here generalise this standard approach and extend the dynamics to charge-separated exciton states using the Merrifield model~\cite{Binder2013}. In this formalism the states of the single-exciton manifold are here given by the product of the localised single-electron and single-hole bases,
\begin{equation}
\label{eq:merrifield_basis}
    \mathcal{B}:=\{\ket{jk}\}_{j,h=1}^N,
\end{equation}
where $\ket{jk}:=\ket{j}_e\otimes\ket{k}_h$, and where $\ket{j}_e$ ($\ket{k}_h$) represents the state of the electron (hole) localised on monomer $j$ ($k$) of a polymeric material.
Every monomer $k$ is also associated with its centre's position $\vec{r}_k:= (x_k,y_k,z_k)$ and a set of local conformational coordinates $\bm{Q}_k$, such as torsional angles and bond-length alternations, as schematically represented in Fig.~\ref{fig:methodology}. These, together with the bonds between pairs of monomers, define the \textit{morphology} of the medium and determine the structure of the exciton's Hamiltonian, as described in the next section.
\subsection{Exciton Hamiltonian}
\label{ss:exciton_hamiltonian}

\noindent
The exciton Hamiltonian $H_\ex= H_0 + H_\F + H_\CT$ is given by the Coulomb term $H_0$, which defines the energetic landscape associated with electron-hole configurations on the polymeric material, and the Frenkel $H_F$ and charge $H_{CT}$ transfer terms, which model the coherent transport of energy and charges, respectively. 
In the electron-hole basis of Eq.~\eqref{eq:merrifield_basis}, the Coulomb term $H_0$ reads
\begin{equation}
    \label{eq:coulomb_term}
    \braket{eh|H_0|e'h'} = \delta_{ee'}\delta_{hh'} \bigg( E_0 - \frac{1}{4\pi \epsilon_0\epsilon_r d_{eh}} \bigg),
\end{equation}
where $E_0^{-1} = (4\pi\epsilon_0\epsilon_r r_0)$ is the exciton binding energy of electron-hole pairs that are localised within the same monomers, i.e., Frenkel excitons. The electron-hole distance, $d_{eh} = r_0 + \lVert \vec{r}_e - \vec{r}_h \rVert$, is obtained from the distance between two monomers offset by the intrinsic electron-hole distance $r_0$ for Frenkel excitons in the considered polymer~\cite{Binder2013}.
The Coulomb term is diagonal in the electron-hole basis, and vanishes for Frenkel excitons. The morphology of the medium introduces static disorder in this term via the inter-monomer distance $d_{eh}$. 

The Frenkel term $H_\F$ models the coherent transport of Frenkel excitons and is given by
\begin{equation}
    \label{eq:frenkel_transfer}
    \braket{eh|H_\F|e'h'} = \delta_{eh}\delta_{e'h'} V_\F(\vec{r}_e,\bm{Q}_e;\vec{r}_{e'},\bm{Q}_{e'}),
\end{equation}
where $V_\F(\vec{r}_j,\bm{Q}_j;\vec{r}_k,\bm{Q}_k)$ is the strength of the Frenkel coupling between a pair of monomers $j,k$, as a function of the morphology of the medium. Frenkel transfer terms preserve charge separation, and can only mediate the transport of Frenkel exciton states, which form a $N$ dimensional subset of the full $N^2$ dimensional Hilbert space associated with the single-exciton manifold of the Merrifield model.

Similarly, the charge transfer (CT) term $H_\CT$ models the coherent transport of individual charges and is given by
\begin{equation}
    \label{eq:charge_transfer}
    \begin{split}
    \braket{eh|H_\CT|e'h'} = & \:\: \delta_{ee'} V_\CT(\vec{r}_h,\bm{Q}_h;\vec{r}_{h'},\bm{Q}_{h'}) + \\  
    &+ \delta_{hh'} V_\CT(\vec{r}_e,\bm{Q}_e;\vec{r}_{e'},\bm{Q}_{e'}),
    \end{split}
\end{equation}
where $V_\CT(\vec{r}_j,\bm{Q}_j;\vec{r}_k,\bm{Q}_k)$ represents the strength of the CT coupling between a pair $j,k$ of monomers, as a function of the morphology of the medium. The CT term can map Frenkel states into charge-separated states and vice versa, and its addition to the Hamiltonian allows exciton states to explore the entirety of the single-exciton manifold.

The dependence of the coupling strengths
$V_\F$ and $V_\CT$ on the morphology of the medium is described in terms of the relative arrangement of pairs of monomers that are conjugated, or close enough to each other to allow for weaker through-space couplings. Without loss of generality, $V_\F$ and $V_\CT$ are assumed to be symmetric in the pairs' indices and to vanish for monomers that are not connected by a bond, or out of range for through-space interactions. An example of the dependence of $V_\F$ and $V_\CT$ on the morphology for PPV-like conjugated polymers is given by Eqs.~\eqref{eq:frenkel_OPV} and~\eqref{eq:ct_OPV} in Sec.~\ref{s:results}. 

\subsection{Exciton-phonon interaction}
\label{ss:coupling_with_environment}
\noindent
Exciton-phonon interactions induce decoherence and relax the exciton states until a steady state is reached. 
The composite dynamics of exciton and phonons is governed by the system-environment Hamiltonian $H=H_\ex+H_\exph+H_\ph$. 
The vibrational modes involved in this process, such as bond-length alternations and ring-breathing modes~\cite{Qin2013,Ma2014,Lyskov2019}, are modelled as a bath of local and independent harmonic oscillators~\cite{Jang2018,Lyskov2019}. 
Let us write the exciton-phonon interaction Hamiltonian $H_\exph$ as 
\begin{equation}
\label{eq:exciton-phonon_hamiltonian}
    H_\exph=\sum_l A_l \otimes B_l,
\end{equation}
such that each coupling operator $A_l$ acting on the exciton space is associated with a bath operator 
\begin{equation}
    B_l = \sum_m g_{l,m} \Big(b_{l,m}^\dagger+b_{l,m}\Big)
\end{equation}
given by a sum over the displacement operators associated with the modes with frequencies $\omega_{l,m}$, written in terms of creation and annihilation operators. The tensor $g_{l,m}$ represents the coupling strengths between the modes $\omega_{l,m}$ and the exciton coupling operators $A_l$. These are assumed to be \textit{weak}, with respect to the characteristic energies of $H_\ex$ and $H_\ph$.
In this notation the phonon Hamiltonian reads
\begin{equation}
    H_\ph =\sum_{l,m} \hbar \: \omega_{l,m} \bigg( b^{\dagger}_{l,m}b_{l,m}+\frac{\mathbb{1}}{2}\bigg).
\end{equation}

The exciton coupling operators $A_l$ model the exchange of energy between the exciton and the vibrational modes. These can either involve a single monomer, or a bond between two monomers. For example, the interaction between a Frenkel state $\ketbra{kk}{kk}$ and ring-breathing modes concerns an individual monomer, here labelled by $k$. Conversely, the interaction between Frenkel transfer terms $\ketbra{jj}{kk}+h.c.$ and bond-length alternation modes is associated with the bond between a pair $j,k$ of monomers. The set of exciton coupling operators $A_l$ considered in this model is given in Table~\ref{tab:couplings}. 


\subsection{Master equation}
\label{ss:master_equation}

\noindent
The dynamics of the reduced exciton state $\rho_t$ is governed by a Lindblad master equation~\footnote{Also known as Gorini–Kossakowski–Sudarshan–Lindblad (GKSL) master equation~\cite{Milz2017}.} that is obtained from the system-environment Hamiltonian $H$ by tracing over the phonon environment~\cite{Lidar2001}
\begin{equation}
\label{eq:master_equation}
\begin{split}
\dot{\rho}_t =&-\frac{i}{\hbar}[H_\ex,\rho_t]
+\sum_{\omega,l} \gamma_{l}(\omega) \bigg(A_l(\omega)\rho_t
A_{l}^{\dagger}(\omega)+\\
&+\frac{1}{2}\Big\{A^{
\dagger}_{l}(\omega)A_l(\omega),\rho_t\Big\}\bigg).
\end{split}
\end{equation}
The Lindblad collapse operators $A_l(\omega)$ are here given by the spectral representation of the coupling operators $A_l$,
\begin{equation}
    \label{eq:lidblad_operators}
    A_l(\omega) = \sum_{E_{\lambda'}-E_\lambda = \omega}  \ketbra{E_\lambda}{E_\lambda} A_l \ketbra{E_{\lambda'}}{E_{\lambda'}},
\end{equation}
where $\ket{E_\lambda}$ are the eigenstates of the exciton Hamiltonian, $H_{\ex}\ket{E_\lambda} = E_\lambda \ket{E_\lambda}$,
such that $\omega$ matches the Bohr frequencies $E_{\lambda'}-E_\lambda$. 
The relaxation rates $\gamma_{l}(\omega)$ are obtained from the Fourier transform of the bath correlation functions $\braket{B_l(t)B_{l'}(t')}=\tr[B_l(t)B_{l'}(t')\rho_\ph]$~\cite{Lidar2001,Heinz-PeterBreuer2002}. Here, $\rho_\ph = \exp(-\beta H_\ph) \mathcal{Z}^{-1}$ is the steady state of the phonon bath, in thermal equilibrium at inverse temperature $\beta = (k_B T)^{-1}$, with partition function $\mathcal{Z} = \tr[\exp(-\beta H_\ph)]$~\cite{Heinz-PeterBreuer2002}.

This general approach, also known as the Bloch-Redfield formalism, allows to obtain relaxation rates from the energy and temperature dependent correlation functions of the bath. The rates obtained this way satisfy thermodynamics and detailed balance condition, leading to the correct thermal equilibrium distribution for long time scales~\cite{Jeske2015}. Here, we obtain a prescription for the dynamics of exciton states in the Markovian approximation, under which condition the master equation can be equivalently expressed in either Bloch-Redfield or Lindblad formalism~\cite{Jeske2015}.

In the Markovian limit that characterises Eq.~\eqref{eq:master_equation}, the bath correlation functions only depend on the time interval $\tau$ between two time steps $t-t'=\tau$~\cite{Lidar2001,Heinz-PeterBreuer2002}. Moreover, in virtue of the mutual independence of the vibrational modes for each pair of coupling operators $A_l$, $A_{l'}$, the bath correlation functions reduce to
\begin{equation}
    \label{eq:bath_correlation_functions}
    \begin{split}
    \braket{B_l(\tau)B_{l'}(0)} &= \delta_{ll'} \sum_m g^2_{l,m}\bigg(\Big(1+\nu_\beta\big(\omega_{l,m}\big)\Big)e^{-i\omega_{l,m} \tau} +\\
    +&\nu_\beta\big(\omega_{l,m}\big)e^{i\omega_{l,m} \tau}
    \bigg),
    \end{split}
\end{equation}
where $\nu_\beta(\omega) = (\exp(\hbar\beta \omega)-1)^{-1}$ is the bosonic distribution function at inverse temperature $\beta$~\cite{Lidar2001,Heinz-PeterBreuer2002}.

To calculate the rates $\gamma_{l}(\omega)$, the sum over the couplings $g_{l,m}$ of Eq.~\eqref{eq:bath_correlation_functions} is replaced with an integral in the frequency domain, $\sum_m g_{l,m}^2\to\int J_l(\omega) d\omega$, where $J_l(\omega)$ is a spectral density that can be used to sample the coupling strengths $g_{l,m}$. To do so we use an Ohmic spectral density $J_l(\omega)$
\begin{equation}
    \label{eq:ohmic-super-ohmic_spectral_density}
    J_l(\omega) = \frac{\Lambda_l}{\hbar}\frac{\omega}{\Omega_l}\exp\bigg(-\frac{\omega}{\Omega_l}\bigg),
\end{equation}
where $\Lambda_l = \hbar \int_0^\infty d\omega J_l(\omega)/\omega$ is the reorganisation energy associated with coupling operator $A_l$~\cite{Lidar2001,Mohseni2008}. In this way the relaxation rates read
\begin{equation}
    \label{eq:lindblad_rates}
    \gamma_{l}(\omega) = 2\pi\bigg(J_l(\omega)(1+\nu_\beta(\omega)) + J_l(-\omega)\nu_\beta(-\omega)\bigg).
\end{equation}
\begin{table}[t]
\begin{tcolorbox}[tabulars*={\renewcommand\arraystretch{1.2}}%
{@{\extracolsep{\fill}\hspace{5mm}}l||l|l@{\hspace{5mm}}},,
boxrule=0.5pt,adjusted title=flush left,halign title=flush left,title = \textbf{Exciton coupling operators} ($A_l$)]
  {} & Site $k$  & Bonds $j,k$\\ \hline\hline
   Frenkel $\;\;\;$ & $\ketbra{kk}{kk}$ & $\ketbra{jj}{kk}+h.c.$ \\
   \hline
   Electron $\;\;\;$ & $\ketbra{k}{k}_e\otimes\mathbb{1}_h$ & $(\ketbra{j}{k}_e+h.c.)\otimes\mathbb{1}_h$ \\
   \hline
   Hole  $\;\;\;$ & $\mathbb{1}_e\otimes\ketbra{k}{k}_h$ $\;\;\;$ & $\mathbb{1}_e\otimes(\ketbra{j}{k}_e+h.c.)$ 
\end{tcolorbox}
\caption{Exciton coupling operators $A_l$ classified by their action on Frenkel, electron, and hole states and transfer terms.}
\label{tab:couplings}
\end{table}

Exciton-phonon interactions also introduce further disorder, which is captured by energy fluctuations $\delta E_l$ in the exciton Hamiltonian, $H_\ex\to H_\ex+\sum_l \delta E_l \:A_l$~\cite{Lee2015,Balzer2021}. These are sampled from a normal distribution with zero average and standard deviation $\sigma_l(\beta)$, given by
\begin{equation}
    \label{eq:static_disorder_standard_deviation}
    \sigma^2_{l}(\beta) = \int_{0}^\infty \:d\omega \:\hbar\: J_{l} (\omega) \coth\bigg(\frac{\hbar \beta \omega}{2}\bigg),
\end{equation}
as done in Ref.~\cite{Sanchez-Carrera2010,Xie2019}. The addition of such energy fluctuations affects site energies and coherent transfer terms, leading to asymmetry in the charge separation landscape and promoting localisation in the centre of mass dyamics~\cite{Walschaers2013,Walschaers2016,Dittel2018}. 

In the next section we use Eq.~\eqref{eq:master_equation} to study exciton transport in conjugated polymers across different morphologies. 
Before discussing the results, let us make a few remarks on the approximations that characterise Eq.~\eqref{eq:master_equation}. First of all, it is worth pointing out that, in Eq.~\eqref{eq:master_equation} we have neglected the Lamb-Shift Hamiltonian term. This term accounts for the contributions of the coupling operators $A_l$ to the coherent dynamics of the exciton. Lamb-Shift corrections have often been neglected in exciton transport problems, especially for the case of local and uncorrelated phonon baths~\cite{Mohseni2008,Caruso2009,Caruso2010}. When applying Eq.~\eqref{eq:master_equation} in Sec.~\ref{s:results} we will assume the exciton Hamiltonian to already include energy shift arising from the weak coupling with the phonon environment. There, Lamb-Shift corrections are expected to be well below the precision limit of a qualitative study focused on order-of-magnitude differences in exciton energies and electronic couplings.

Another important observation is that we have assumed the vibrational modes to be uncoupled with each other, which is the reason why Eq.~\eqref{eq:master_equation} does not feature combinations of $A_l(\omega)$ and $A_{l'}(\omega)$ Lindblad operators. It is reasonable to imagine that this approximation would not be justified for some combinations of coupling operators, such as for those vibrational modes coupled with the same monomer. When spatial correlations become important, e.g., when phonon modes are delocalised over several monomers, they can be simply accounted for and included into Eq.~\eqref{eq:master_equation} via the addition of terms in $A_l$ and $A_{l'}$, and the associated rates arising from the spatial dependence of the correlation functions $\langle B_l(t),B_{l'}(0) \rangle$~\cite{Jeske2013a}. The presence of strong and long-range spatial correlations allows for the formation of decoherence free subspaces and gives rise to super and subradiance phenomena~\cite{Batey2015}, which have been studied for biological light-harvesting complexes~\cite{Jeske2015}, but remain rather unexplored for the case of organic polymers.

Finally, the weak coupling approximation for the exciton-phonon interaction, necessary to obtain Eq.~\eqref{eq:master_equation}, is arguably the main source of deviations from a rigorous description of the exciton transport. While some vibrational modes, such as those associated with torsional angles and foldings of a polymer chain, are very likely to be well described by a weak coupling approximation, others modes, such as bond-length alternations, can couple strongly with the excitons. In such cases it is possible to overcome the limitations imposed by Eq.~\eqref{eq:master_equation} by performing a polaron-transformation~\cite{Grover1971,Pouthier2006,Jang2008,McCutcheon2011}, which models the dynamics of an exciton followed by nuclear deformations, i.e., a polaron (or \emph{dressed-exciton}), using master equations similar to Eq.~\eqref{eq:master_equation}, such as the secular polaron-transformed
Redfield equation (sPTRE)~\cite{Athanasopoulos2007,Coropceanu2007b,Ortmann2009,Fishchuk2013,Lee2015,Wilner2015,Binder2018,Takahashi2019,Balzer2021}.
\begin{table}[t]
\begin{tcolorbox}[tabulars*={\renewcommand\arraystretch{1.2}}%
{@{\extracolsep{\fill}\hspace{5mm}}ll|l@{\hspace{5mm}}},adjusted title=flush left,halign title=flush left,
boxrule=0.5pt,title = {\textbf{ Exciton Hamiltonian parameters} --- OPV, PPV}]
\textit{Parameter} & \textit{} & \textit{Hartree AU} \\
\hline\hline
Frenkel electron-hole distance & $r_0$ & $12.7$ \\ \hline
Exciton biding energy & $E_0$ $\;$ & $7.90\times10^{-2}$ \\ \hline
Frenkel transfer parameters  & $j_0 $ & $5.03\times10^{-3}$ \\ \hline
{} & $j_2 $ & $1.20\times10^{-2}$ \\ \hline
{} & $j_4 $ & $3.23\times10^{-2}$ \\ \hline
Charge transfer parameter  & $t_0 $ & $9.78\times10^{-2}$ \\ \hline
Relative permittivity  & $\epsilon_0 $ & $1$ 
\end{tcolorbox}
    \caption{Exciton Hamiltonian parameters used in Eqs.~\eqref{eq:coulomb_term},~\eqref{eq:frenkel_OPV}, and~\eqref{eq:ct_OPV} for the OPV and PPV-like materials considered in Secs.~\ref{ss:opv_hexamer} and~\ref{ss:ppv_polymer},  expressed in Hartree atomic units~\cite{Binder2013}. Frenkel transfer parameters have been scaled in Sec.~\ref{ss:tangle_octamers} in order to study the dependence of transport properties on the strength of the electronic couplings.}
    \label{tab:constants_OPV}
\end{table}

\section{Results}
\label{s:results}
\noindent
We now use Eq.~\eqref{eq:master_equation} to study room-temperature exciton transport in some representative conjugated oligomers and polymers, exploring the transition from ordered to disordered morphologies. To compare our findings with previous results, we base our calculation upon the electronic properties of OPV and PPV, used in Refs.~\cite{Binder2013,Lyskov2019}.
In particular, we focus on the following systems. First, we consider a short OPV hexamer to explore centre of mass (CoM) and charge separation (CS) dynamics with a full Merrifield exciton approach, in Sec.~\ref{ss:opv_hexamer}. Then, in Sec.~\ref{ss:ppv_polymer}, we model a long PPV polymer with 51 repeated unit to explore intrachain Frenkel exciton dynamics across different morphological regimes. Finally, we examine the role of intrachain and interchain transport and its dependence on electronic coupling strength in Sec.~\ref{ss:tangle_octamers}, by looking at Frenkel exciton transport in OPV oligomers.

The different regimes of disorder, from \textit{crystalline} to \textit{amorphous}, are here achieved sampling torsional angles and bond length-alternations from a normal distribution with zero average and tunable standard deviation. 
The functional dependence of Frenkel and CT terms on the relative arrangement between two conjugated monomers, given by $V_\F(d,\theta)=f(d)\Theta_\F(\theta)$ and $V_\CT(d,\theta)=f(d)\Theta_\CT(\theta)$, respectively, is calculated upon their distance $d$ and the relative angle $\theta$ between the two monomers' planes. The angular dependence $\Theta_\F(\theta)$ and $\Theta_\CT(\theta)$ is modelled as in Ref.~\cite{Binder2013},
\begin{align}
    \label{eq:frenkel_OPV}
    &\Theta_F(\theta) = (j_4-j_0) \cos^4\theta + (j_2-j_0) \cos^2\theta - 2j_0, \\
    \label{eq:ct_OPV}
    &\Theta_\CT(\theta) = t_0\cos\theta,
\end{align}
while the dependence on the intra-monomer distance is heuristically modelled with a Gaussian factor $f(d)=\exp[(d-r_0)^2/4r_0^2]$. 

The parameters used in Eqs.~\eqref{eq:coulomb_term},~\eqref{eq:frenkel_OPV}, and~\eqref{eq:ct_OPV} are given in Tab.~\ref{tab:constants_OPV} and expressed in Hartree atomic units, while the relative permittivity of the considered materials is assumed to be $\epsilon_r=1$ for simplicity. The characteristic electronic couplings associated with Frenkel and charge-transfer terms can be as high as 2 eV, and, thus, much larger than the 26 meV associated with the thermal energy of the phonon bath at room temperature of 300K.
When required, Frenkel through-space couplings are included for non-conjugated monomers that are separated by $\approx r_0$, with characteristic energies around 10 meV~~\cite{Nelson2017,Lyskov2019}.

\begin{figure}[t]
    \centering
    \includegraphics[width=0.49\textwidth]{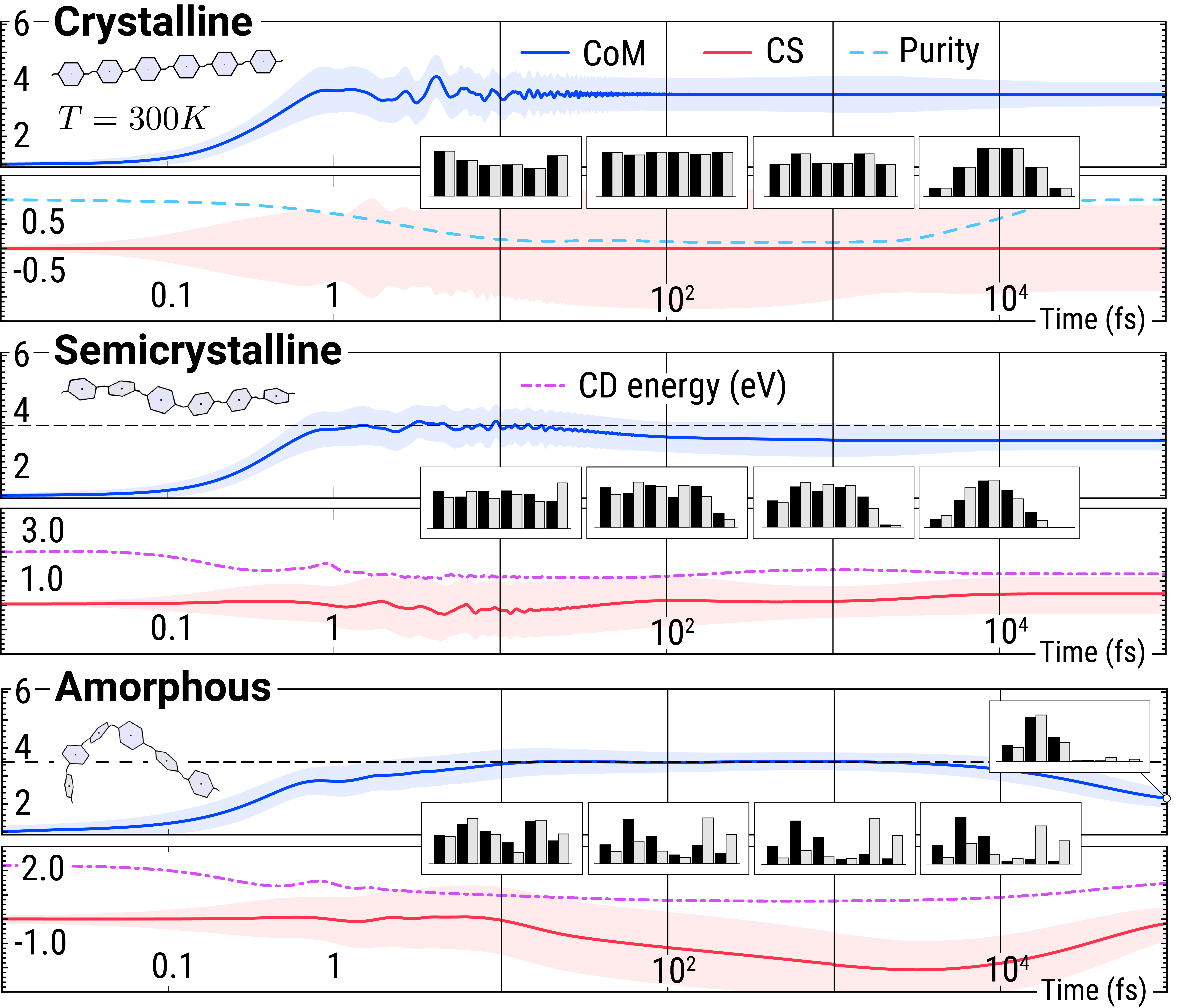}
    \caption{Exciton transport for different instances of OPV hexamer morphologies at room temperature $T = 300 K$. For each case the initial state $\rho_0$ is a Frenkel exciton localised on the first monomer. Expectation value (solid lines) and standard deviation (shaded areas) of CoM (blue) and CS (red) are expressed in the sites basis. The energy required  for charge dissociation (dotted-dashed magenta line) is shown in eV (on the same $y$-axis as CS). The histograms insets schematically represent the populations (not in scale) of the partial states of electron (black bars) and hole (light gray bars) on each site at a given time.}
    \label{fig:OPV_hexamer}
\end{figure}

\begin{figure*}[ht]
    \centering
    \includegraphics[width=\textwidth]{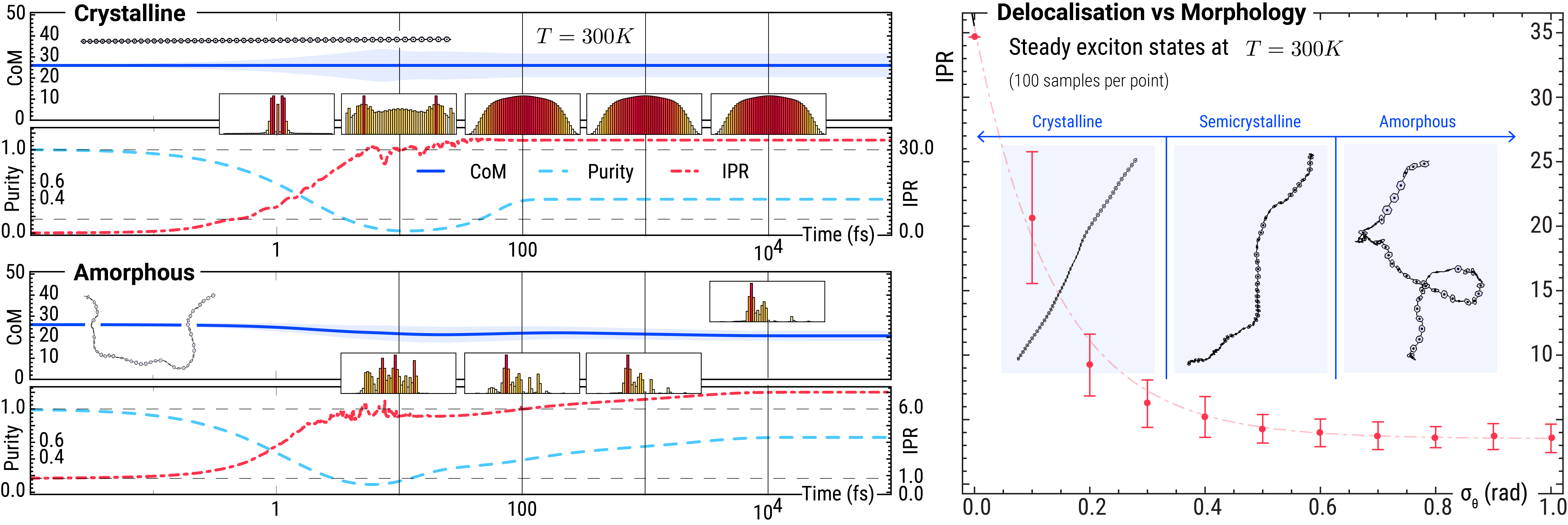}
    \caption{(\emph{Left}) Exciton transport for crystalline (\emph{top-left}) and amorphous (\emph{bottom-left}) PPV polymers with 51 monomers, at room temperature $T = 300 K$. For each case the initial state $\rho_0$ is a Frenkel exciton localised on the central monomer ($k=26)$. Expectation value (solid lines) and standard deviation (shaded areas) of CoM (blue) are expressed in the sites basis. Delocalisation is monitored by purity (dashed light blue lines) and average $\texttt{IPR}$ in the sites basis (dashed-dotted red line). The histograms insets schematically represent the populations (not in scale) of the exciton on each site at a given time. (\emph{Right}) The dependence of steady state delocalisation on the morphology is studied varying the standard deviation $\sigma_\theta$ of random torsional angles, and sampling 100 PPV polymers with 51 repeated units for each value of $\sigma_\theta$.}
    \label{fig:ppv_51}
\end{figure*}

 Reorganisation energies $\Lambda_l$ and cut-off frequencies $\Omega_l$ associated with vibrational modes define relaxation rates $\gamma_l(\omega)$ and energy fluctuations $\sigma_l(\beta)$, as prescribed by Eq.~\eqref{eq:ohmic-super-ohmic_spectral_density}~\cite{Lidar2001,Heinz-PeterBreuer2002}. For the considered systems we have chosen reorganisation energies of 500 meV and 50 meV for monomer-local and bond-local vibrational modes, respectively. The chosen cut-off frequencies are 1500 $\text{cm}^{-1}$ for modes that couple with Frenkel states, 1000 $\text{cm}^{-1}$ for modes that couple with individual charges and Frankel transfer terms, and 500 $\text{cm}^{-1}$ for modes that couple with CT terms~\cite{Lefrant1989,Dykstra2009,Oberhofer2017,Lyskov2019}.

To propagate an initial exciton state $\rho_0$ we exactly solve the system of linear differential equations associated with Eq.~\eqref{eq:master_equation} written in Liouville space~\cite{Havel2003}. In this way we obtain a map $\Lambda_t$ for the dynamics of the state $\rho_t=\Lambda_t[\rho_0]$. Exciton transport properties are studied evaluating time-dependent expectation value and standard deviation of CoM and CS operators. These can be expressed in terms of monomer indices $\braket{eh|X_{\text{CoM}}|e'h'} = \delta_{ee'}\delta_{hh'}(e+h)/2$, and $\braket{eh|X_{\text{CS}}|e'h'} = \delta_{ee'}\delta_{hh'}(e-h)$, as done in Ref.~\cite{Binder2013}. Alternatively, they can be expressed in terms of monomer coordinates $\vec{r}_k = (x_k,y_k,z_k)$, by replacing $e$ ($h$) with a given Cartesian coordinate, such as $x_{e}$ ($x_{h}$).

Localisation properties of Frenkel states are inferred from \textit{purity}, $\mathcal{P}[\rho_t]=\tr[\rho_t^2]$, and average \textit{inverse participation ratio} ($\texttt{IPR}$)~\cite{ Cho2005,Dykstra2009,Kranz2016a,Balzer2021,Scholes2019}. Let $\rho = \sum_{k}p_k \ketbra{r_k}{r_k}$ be expressed in its eigenbasis $\{\ket{r_k}\}$, and let $\{\ket{n}\}$ be the sites basis then
\begin{equation}
    \label{eq:gIPR}
    \overline{\texttt{IPR}}(\rho) =  \sum_k p_k \bigg( \sum_{n} |\braket{n|r_k}|^4\bigg)^{-1},
\end{equation}
i.e., the weighted sum of the $\texttt{IPR}=1/\sum_n |\braket{n|\psi}|^4$ for each pure state $\ket{r_k}$ making up the density operator $\rho$~\cite{Moix2013,Balzer2021}. From the $\texttt{IPR}$, delocalisation length can be estimated using $l = (\overline{\texttt{IPR}}(\rho))^{1/d}$ for a $d$-dimensional system~\cite{Balzer2021}. Since the considered systems are either one-dimensional or ensembles of coupled one-dimensional systems we estimate the delocalisation length simply using $\overline{\texttt{IPR}}(\rho)$.

\subsection{OPV hexamer}
\label{ss:opv_hexamer}
\noindent
This system consists of six conjugated monomers and it is studied with a full Merrifield exciton approach, i.e., including all the possible CS states. For each morphology, we initialise the exciton in a Frenkel state localised on the \textit{first} monomer. For comparison, a similar system was studied in Ref.~\cite{Binder2013} with a MCTDH method, requiring around $10^6$ configurations for its numerical implementation, as opposed to the 36 required for the solution of Eq.~\eqref{eq:master_equation}~\footnote{As discussed in Sec.~\ref{s:methodology} this advantage in terms of computational cost comes at the expense of several approximations that reduce the accuracy of the master equation.}

The crystalline morphology, shown in Fig.~\ref{fig:OPV_hexamer} (\emph{top}), is obtained for a perfectly planar arrangement of the monomers with constant bond-length alternation and no static disorder. It is characterised by trivial CS dynamics, which remains constant and equal to zero for all times. 
The exciton undergoes a initial ultrafast ballistic transport regime (1-10 fs), followed by a diffusive transport regime that lasts for the first 100 fs. During the evolution the partial states of electron and hole are perfectly symmetric, as shown in the histograms insets of Fig.~\ref{fig:OPV_hexamer} representing the population (not in scale) of the partial states of electron (black bars) and hole (light gray bars) on each site at a given time.

The semicrystalline morphology, an instance of which is shown in Fig.~\ref{fig:OPV_hexamer} (\emph{middle}), is obtained for torsional angles normally distributed with standard deviation of 0.5 rad. The presence of disorder allows for a non trivial CS dynamics, thus reducing the energy barrier required for charge dissociation (CD) of the exciton. Despite the CoM dynamics has similar features to those of the crystalline morphology, asymmetry in the partial states of electron and hole is generally present both during the evolution and in the steady states. 

The amorphous phase, an instance of which is shown in Fig.~\ref{fig:OPV_hexamer} (\emph{bottom}), is obtained for torsional disorder with standard deviation of 1 rad. It is characterised by noticeable CS dynamics, with electron-hole separation over more than 2 monomers, lasting for hundreds of femtoseconds. Such CS dynamics can considerably reduce the CD energy for relatively long time spans of hundreds of femtoseconds. Remarkably, high-disorder allows for the formation of long-lived non-equilibrium states, before thermal equilibrium is reached around 100 ps. Exciton lifetimes may be shorter than the time required to reach thermal equilibrium in such systems. This implies that thermal equilibrium states of the excitons are generally not experimentally accessible.
\begin{figure*}[t]
    \centering
    \includegraphics[width=\textwidth]{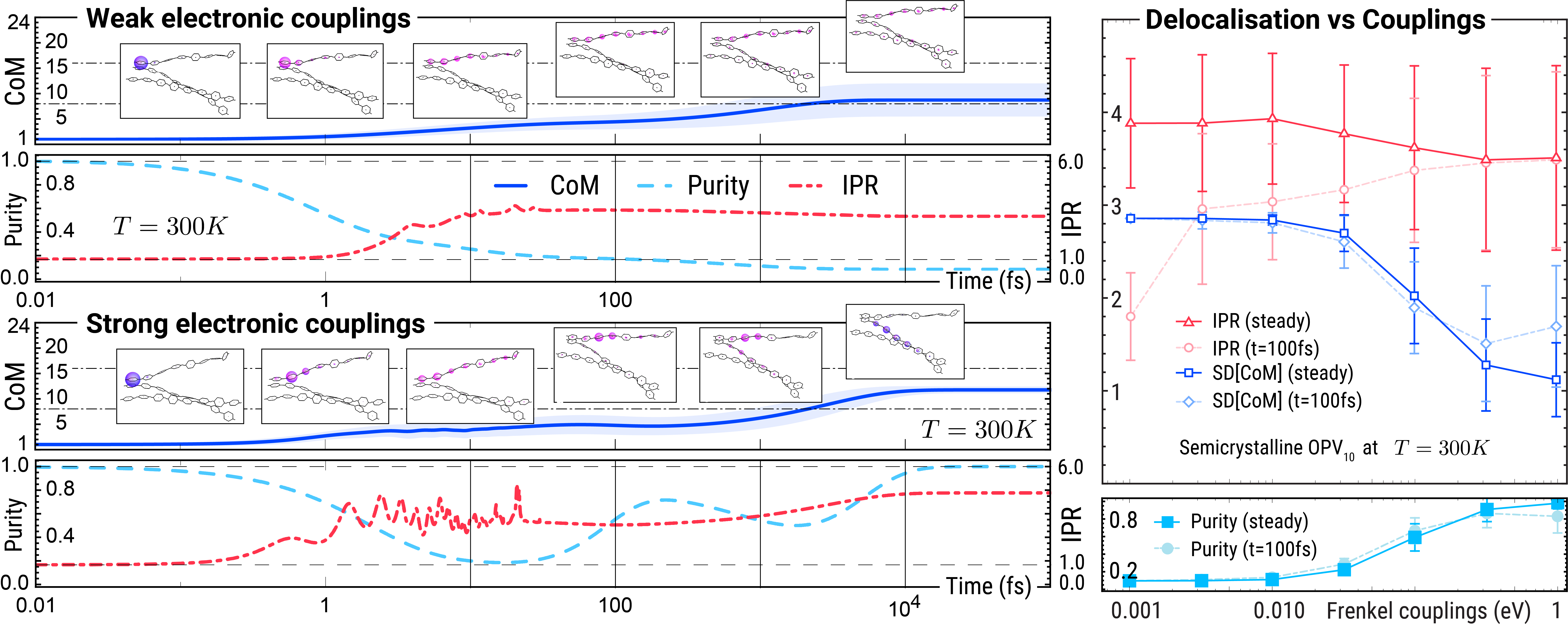}
    \caption{(\emph{Left}) Exciton transport for an arrangement of three semicrystalline OPV octamers with weak (10 -- 100 meV) (\emph{top-left}) and strong (0.1 -- 1 eV) (\emph{bottom-left}) electronic couplings at room temperature $T = 300 K$. For each case the initial state $\rho_0$ is a Frenkel exciton localised on first monomer of the first octamer. Expectation value (solid lines) and standard deviation (shaded areas) of CoM (blue) are expressed in the sites basis. Delocalisation is monitored by purity (dashed light blue lines) and average $\texttt{IPR}$ in the sites basis (dashed-dotted red line). The insets schematically represent the populations of the exciton on each site at a given time. (\emph{Right}) The dependence of exciton delocalisation on the electronic couplings is studied varying the characteristic strength of Frenkel couplings from 10 meV to 1 eV. For each coupling regime we sampled 100 OPV oligomers with 10 repeated units and semicrystalline morphology ($\sigma_\theta = 0.5$ rad). Delocalisation is studied by evaluating $\texttt{IPR}$, purity and standard deviation of CoM for steady states and for states at 100 fs, with $\rho_0$ initialised in a Frenkel state localised on the first monomer.}
    \label{fig:tangle}
\end{figure*}

\subsection{PPV polymer}
\label{ss:ppv_polymer}
\noindent

\noindent
We now consider a PPV polymer with 51 repeated units. We restrict the dynamics to the Frenkel manifold to study the transport regimes of the CoM while limiting the computational cost. Excitons are here initialised in a Frenkel state localised on the \textit{central monomer} ($k=26$) of each considered polymer.

The crystalline morphology, shown in Fig.~\ref{fig:ppv_51} (\emph{top-left}), is characterised by a transition from 
ballistic to diffusive exciton transport during the first 100 fs, in analogy with the results of Ref.~\cite{Lyskov2019}. The remaining part of the dynamics consists in a relaxation process that guides the exciton towards thermal equilibrium. 
The absence of disorder in the sites energies and electronic couplings allows for the formation of highly delocalised steady states, with $\overline{\texttt{IPR}} > 30$. 

However, in contrast with the results of Ref.~\cite{Lyskov2019}, steady state populations are not evenly spread across the polymer, and thus only partially mixed. This is because the decoherence model used in Ref.~\cite{Lyskov2019} does not account for the spectral response between exciton coupling operators and the phonon bath. The results of Ref.~\cite{Lyskov2019} can be qualitatively reproduced using Eq.~\eqref{eq:master_equation} by replacing the Lindblad operators $A_l(\omega)$ with the coupling operators $A_l$, and the associated relaxation rates $\gamma_l(\omega)$ with frequency independent rates $\gamma_l$. In such simplified limit, steady states become diagonal in the sites' basis. Here, instead, we use a Bloch-Redfield approach for rates and dissipators and obtain the correct thermal equilibrium states, which are diagonal in the exciton Hamiltonian basis.

As disorder is increased, localisation becomes more evident, undermining the ultrafast ballistic transport regime that would otherwise characterise a crystalline morphology. The amorphous morphology, an instance of which is shown in Fig.~\ref{fig:ppv_51} (\emph{bottom-left}), is characterised by sub-diffusive exciton transport within the first 100 fs, followed by thermal relaxation. States are remarkably less delocalised both during the dynamics and at equilibrium, with $\overline{\texttt{IPR}}\approx6$ and CoM standard deviations much lower than for the crystalline morphology.

To further examine the relation between morphology and exciton delocalisation we study the $\texttt{IPR}$ of steady states at room temperature for different morphological regimes. To do so we vary the standard deviation $\sigma_\theta$ for the random torsional angles between 0 (crystalline) and 1 rad (amorphous), sampling 100 different PPV polymers with 51 monomers for each morphology. The results, presented in Fig.~\ref{fig:ppv_51} (\emph{right}), show the high sensitivity of exciton delocalisation to conformational defects. Delocalisation rapidly drops from $\overline{\texttt{IPR}}\approx 35$ for crystalline polymers to $\overline{\texttt{IPR}}\approx 4$ for amorphous ones.

\subsection{Dependence on electronic couplings}
\label{ss:tangle_octamers}
\noindent
We now study the dynamics of Frenkel excitons for fixed morphology while varying the strength of the electronic couplings. This allows us to explore the different regimes of exciton transport and thermal equilibrium that are associated with weak (10 -- 100  meV) or strong (0.1 -- 1 eV) electronic couplings. 
First, we illustrate such difference by looking at Frenkel exciton dynamics for a system given by three closely arranged OPV octamers that interact via through-space couplings, shown in Fig.~\ref{fig:tangle} (\emph{left}).

Systems characterised by weak electronic couplings rapidly lose their coherence, which vanishes within the first 10 -- 20 fs. As shown in Fig.~\ref{fig:tangle} (\emph{top-left}), exciton transport is incoherent both during an initial intrachain transport over one oligomer and during the slower interchain transport across the different oligomers. The thermal energy is sufficient to populate several eigenstates of the exciton Hamiltonian. Thermal equilibrium is therefore characterised by low-purity and large CoM standard deviation.

Strong electronic couplings ($\approx 1$ eV) also display ultrafast intrachain dynamics, followed by a slower interchain transport. However, the thermal energy is not high enough to populate several states of the exciton Hamiltonian, therefore the steady states are rather pure and localised around the most energetically favourable clusters of conjugated monomers, as shown in Fig.~\ref{fig:tangle} (\emph{bottom-left}). 

To systematically explore the dependence of exciton delocalisation on the strength of the electronic couplings (and thus on the thermal energy) we study $\texttt{IPR}$, purity and CoM standard deviation of transient (100 fs) and steady states of semicrystalline oligomers. The characteristic strength of Frenkel couplings is varied from 10 meV to 1 eV. For each coupling regime we sample 100 OPV oligomers with 10 repeated units and fixed semicrystalline morphology ($\sigma_\theta = 0.5$ rad). As shown in Fig.~\ref{fig:tangle} (\emph{right}), the average $\texttt{IPR}$ decreases only slightly ($\approx 10\%$) over two orders of magnitude of electronic couplings strength. However, transport properties change dramatically, with CoM standard deviation and purity varying remarkably for both transient and steady states.

\subsection{Final remarks}
\label{ss:final_remarks}
\noindent
Using the model introduced in this article the dynamics of excitons in conjugated polymers is understood as a quantum thermalisation process, whose features strongly depend on the amount of disorder (i.e., morphology) and on the relative magnitude of thermal energy and electronic couplings (i.e., temperature). The non-equilibrium dynamics is characterised by an ultrafast and ballistic transport transient, followed by an intermediate diffusive or sub-diffusive process that occurs both through-bond and through-space. The dynamics culminates with thermal relaxation. Increasing disorder enables localisation, thus decreasing delocalisation lengths and $\texttt{IPR}$ and hindering the efficacy of exciton transport. The amorphous phase is characterised by the presence of long-lived non-equilibrium states and prominent charge-separation dynamics, which however is limited to a few monomers, as previously reported in Ref.~\cite{Qin2013}.

In polymers characterised by weak electronic couplings, of the order 10 -- 100 meV, the environment \textit{heats} the system, driving the excitons low-purity thermal states, with non-negligible populations across the whole spectrum of energy eigenstates of the exciton Hamiltonian. Even though the intrinsic localisation lengths are low due to the presence of disorder, the thermal energy of the phonon bath (around 26 meV at room temperature) is sufficient to populate monomers that are far apart from the initial exciton location. In contrast, in materials characterised by strong Frenkel and charge-transfer couplings ($\approx 1$ eV), the phonon environment \textit{cools} the system, leading the excitons to high-purity thermal state, with low-energy eigenstates of the exciton Hamiltonian being the only few populated ones. 
For this reason, despite the slower intrachain transport, systems with weak electronic couplings can display a more efficient transport mechanism over the picosecond time-scale at room temperature. This becomes particularly important for triplet excitons, which typically have lower mobility and longer lifetime than singlet excitons.

\section{Conclusions}
\label{s:conclusions}
\noindent
In this article we have introduced a general master equation to study the dynamics of Merrifield excitons in conjugated polymers as a function of temperature and morphology of the medium. Using this method we have explored the general features of energy and charge transport in some representative systems based on PPV's electronic properties, confirming well known paradigms of quantum transport in disordered systems, and revealing non-equilibrium features such as long-lived states and charge-separation dynamics. 

Beyond the qualitative understanding of exciton dynamics in amorphous polymers, we expect our method to be applicable for the quantitative study of energy and charge transport properties of specific materials. This can be done using a multi-scale approach based on molecular dynamics and first-principle calculations, as demonstrated in Refs.~\cite{Nakano2019,Xie2019,Lyskov2019}.
We also anticipate that this approach could be used to study disorder-dependent effects in OSCs, such as trapped-charge induced photoluminescence peak displacement~\cite{Bardeen2011,Bolinger2011} and 
individual charge-carrier pathways in amorphous polymers~\cite{Wilma2016}.

\begin{acknowledgments}
\noindent
This research was supported by the Australian Research Council under grant number CE170100026.
FC thanks Igor Lyskov and Ivan Kassal for insightful discussions.
This research was undertaken with the assistance of resources from the National Computational Infrastructure (NCI), which is supported by
the Australian Government. The authors acknowledge the people of the Woi wurrung and Boon wurrung language groups of the eastern Kulin Nations on whose unceded lands we work. We respectfully acknowledge the Traditional Custodians of the lands and waters across Australia
and their Elders: past, present, and emerging.
\end{acknowledgments}

\bibliography{main.bbl}

\end{document}